\documentclass[12pt]{article}
\usepackage{epsfig}
\usepackage{pst-plot,url,epsf}
\newcommand{\beq}{\begin{equation}}
\newcommand{\eeq}{\end{equation}}
\newcommand{\bea}{\begin{eqnarray}}
\newcommand{\eea}{\end{eqnarray}}
\newcommand{\beas}{\begin{eqnarray*}}
\newcommand{\eeas}{\end{eqnarray*}}
\newcommand{\epm}{e^+e^-}

\newcommand{\ra}{\rightarrow}

\newcommand{\eesixf}{e^+ e^- \ra b f_1 \bar{f'_1} \bar{b} f_2 \bar{f'_2}}
\newcommand{\AmS}{{\protect\the\textfont2
  A\kern-.1667em\lower.5ex\hbox{M}\kern-.125emS}}

\begin{document}
\thispagestyle{empty}
\begin{flushright}
November 2009
\end{flushright}

\vspace*{1.5cm}

\begin{center}
{\LARGE\bf Signal and Background in $\epm \ra t \bar t H$\footnote{Presented at the XXXIII International Conference of Theoretical Physics, `` Matter to the Deepest'', Ustro\'n, Poland, September 11--16, 2009.}
}\\
\vspace*{2cm}
%----------------------------
Karol Ko\l odziej\footnote{{\em E-mail:} karol.kolodziej@us.edu.pl}
and Szymon Szczypi\'nski\footnote{{\em E-mail:} simon\_t@poczta.fm}\\[6mm]
{\small\it Institute of Physics, University of Silesia\\ 
ul. Uniwersytecka 4, PL-40007 Katowice, Poland}\\
\vspace*{2.5cm}
{\bf Abstract}\\
\end{center}
We discuss the reaction of associated production of a top quark pair
and a Higgs boson at the future $\epm$ linear collider that can be
used to determine the top--Higgs Yukawa coupling. Taking into account 
decays of the top quarks and the light Higgs boson leads to reactions with 
8 fermions
in the final state which, in the framework of the standard model,
receive contributions typically from many thousands Feynman diagrams. An
overwhelming majority of the diagrams comprises the off resonance background 
to the resonant signal of associated production and decay of the top quark 
pair and Higgs boson. 
We address the signal and background issue by comparing cross sections
calculated with the signal diagrams only and with the complete sets of
the diagrams.

\vfill

\newpage

\section{Introduction}

If the Higgs boson exists in Nature then it will be most probably discovered 
at the LHC. However, its production and decay properties, which are crucial 
for verification of the electroweak (EW) symmetry breaking mechanism, can be 
best studied in a clean environment of the $\epm$ collisions at a linear 
collider.
There are two projects of the linear collider being developed in a world
wide collaboration: the International Linear Collider (ILC)
\cite{ILC}, with the collision energy of 0.5--1~TeV, and the
Compact Linear Collider (CLIC) \cite{CLIC}, with the nominal collision 
energy of 3~TeV.

\begin{figure}[htb]
\vspace{110pt}
\includegraphics{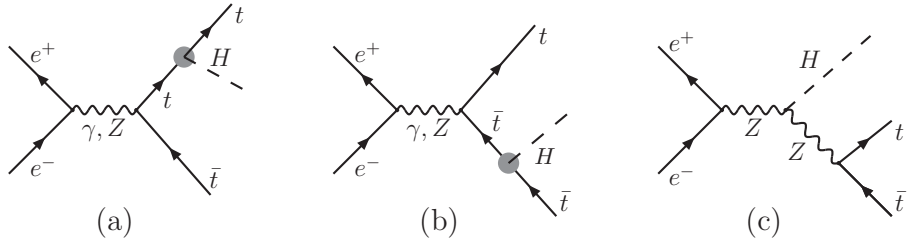}
\caption{Lowest order Feynman diagrams of reaction (\ref{eetth}) in the
unitary gauge.}
\label{fig:eetth}
\end{figure}

The Higgs boson mass constraints from direct searches at LEP 
give a lower limit of $m_H > 114.4$~GeV at 95\% CL, 
the direct searches  at Tevatron 
exclude the standard model (SM) Higgs boson mass in the region 
$160\;{\rm GeV} < m_H < 170$~GeV and the
virtual effects it has on precision EW observables in the framework of SM
strongly favor a light Higgs boson. 
These constraints indicate that the SM Higgs boson should be
searched for in the mass range $114.4\;{\rm GeV} < m_H < 160$~GeV and, if
it has been found, the $t\bar t H$ Yukawa coupling
$g_{ttH}={m_t}/{v}$, with $v=(\sqrt{2}G_F)^{-1/2}\simeq 246 \rm{GeV}$,
can then be best determined in the reaction \cite{eetth}
\bea
\label{eetth}
e^+ e^- \rightarrow t \bar{t} H.
\eea
The lowest order Feynman diagrams of (\ref{eetth}) are shown in 
Fig.~\ref{fig:eetth}, where we have neglected the diagrams with the Higgs boson
coupling to $\epm$ and the $g_{ttH}$ coupling has been indicated by the gray
circle.
The diagrams in Fig.~\ref{fig:eetth}(a) and Fig.~\ref{fig:eetth}(b) 
dominate the cross section of (\ref{eetth}) therefore it is almost 
proportional to $g_{ttH}^2$.

\begin{figure}[htb]
\vspace{110pt}
\includegraphics{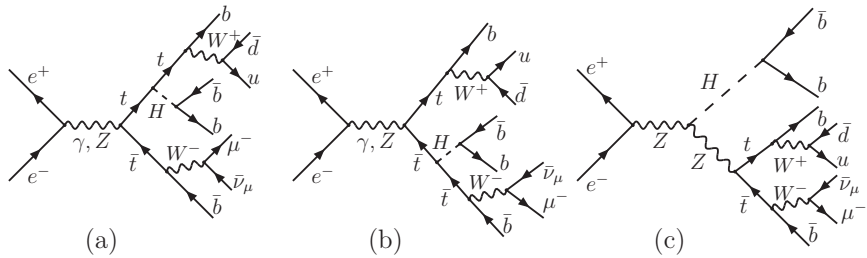}
\caption{Lowest order signal Feynman diagrams of (\ref{semileptonic}).
         The remaining diagrams are obtained by 
permutations of the identical $b$- and $\bar b$-quarks.}
\label{fig:ee8f}
\end{figure}

As $t$ and $\bar t$ decay before they hadronize, 
predominantly into $b W^+$ 
and $\bar{b} W^-$, the $W'$s decay into $f\bar{f'}$-pairs and, 
if $m_{H} < 140$ GeV, the Higgs boson decays dominantly into a $b\bar{b}$-pair,
reaction (\ref{eetth}) is observed through reactions of the form
\bea
\label{ee8f}
\eesixf b\bar b,
\eea
where
$f_1, f'_2 =\nu_{e}, \nu_{\mu}, \nu_{\tau}, u, c$ and
$f'_1, f_2 = e^-, \mu^-, \tau^-, d, s$. Different channels of (\ref{ee8f})
are usually classified according to the decay channels of the $W$ bosons.
For example, the reactions
\bea
\label{leptonic}\epm &\ra& b \bar b b \bar b \tau^+ \nu_{\tau} \mu^- \bar \nu_{\mu},\\
\label{semileptonic}
\epm &\ra& b \bar b b \bar b u\bar d \mu^- \bar \nu_{\mu},\\
\label{hadronic}
\epm &\ra& b \bar b b \bar b c\bar s s \bar c
\eea
represent the leptonic, semileptonic and hadronic detection channels
of (\ref{eetth}). In the the unitary gauge of SM, neglecting the Yukawa 
couplings of the fermions lighter than $c$-quark and $\tau$-lepton,
they receive contributions from $21\,214$, 
$26\,816$ and $240\,966$ lowest order Feynman diagrams, respectively.
An overwhelming majority of the diagrams comprises background 
to the signal of associated production and decay of the top quark pair 
and Higgs boson that in each channel gets contributions from
20 Feynman diagrams which involve 3 resonant propagators 
of the $t$- and $\bar t$-quarks and Higgs boson at a time. For example,
the signal diagrams of (\ref{semileptonic}) are shown
in Fig.~\ref{fig:ee8f}. 

In the present paper, we will address the signal and background issue 
by comparing cross sections calculated with the 20 signal diagrams only, with 
the complete sets of the diagrams and with the neglect of the gluon exchange
contributions. The comparison will be performed first in the presence
of basic particle identification cuts and then with additional invariant
mass cuts that should allow for identification of the top quarks, secondary 
$W$ bosons and Higgs boson.
We will also illustrate pure off mass shell effects
by comparing the signal cross section calculated with the full 8 particle 
kinematics and in the narrow width approximation.

\section{Calculational details}

Because of large numbers of the Feynman diagrams the cross sections of 
reactions (\ref{leptonic}), (\ref{semileptonic}) and (\ref{hadronic}) 
must be computed in a fully automatic way. We will use to this end a recently 
released multipurpose Monte Carlo (MC) program {\tt carlomat} \cite{carlomat}.

The program itself is written in Fortran 90/95 and it
generates Fortran routines for the matrix element in the lowest order
of SM for a user specified process, taking into account both 
the EW and quantum chromodynamics (QCD) contributions. Any kind of particles 
in the initial state, 
including quarks and gluons,  are possible, with up to 12 external particles.
The MC summing over helicities has been implemented in the program as
an option, however, an explicit helicity summing is also possible. When doing 
so, spinors or polarization vectors representing particles, or building
blocks of the Feynman diagrams, are computed only once,
for all the helicities of the external particles they are made of, and stored 
in arrays.
The program generates phase space parametrizations, which take into account 
peaks in every Feynman diagram. They are automatically implemented into
a multichannel MC integration routine for the cross section 
computation and event generation. Weights with which different kinematical 
channels contribute to the cross section are adapted iteratively. It should be
stressed at this point that the phase space integration 
is the most time consuming part of the program.
The initial state radiation has been implemented recently
and some extensions of the SM and effective models are being implemented 
at the moment.

\section{Signal and background}

The results presented in this section have been obtained with the initial 
physical parameters of \cite{KS}. In all tables presented below, the numbers 
in parenthesis show the MC uncertainty of the last decimal.

Let us concentrate on the semileptonic reaction (\ref{semileptonic}).
The easiest way to calculate the signal cross section of associated 
production of the
top quark-pair and Higgs boson is to apply
the narrow width approximation that is given by
\bea
\label{NWA}
\sigma_{\rm signal}^{\rm NWA}=
\sigma (e^+ e^- \rightarrow t \bar{t} H)\times 
\frac{\Gamma_{W^+\rightarrow u \bar{d}}}{\Gamma_{W}}\times 
\frac{\Gamma_{W^-\rightarrow \mu^- \bar{\nu}_{\mu}}}{\Gamma_{W}}
\times\frac{\Gamma_{H\rightarrow b \bar{b}}}{\Gamma_{H}},
\eea
where $\sigma (e^+ e^- \rightarrow t \bar{t} H)$ is the lowest order cross
section of (\ref{eetth}), the $\Gamma$'s are the lowest order widths and
we have assumed $\Gamma_{t \rightarrow b W^+}/\Gamma_{t}=
\Gamma_{\bar t \rightarrow \bar b W^-}/\Gamma_{t}=1$. Then
we calculate the signal cross section $\sigma_{\rm signal}$ once more,
taking into account the 20 signal diagrams
of (\ref{semileptonic}) and applying the full 8-particle kinematics, 
{\it i.e.},
integrating the squared matrix element of (\ref{semileptonic}) over 
20-dimensional phase space. The results for 
$\sigma_{\rm signal}^{\rm NWA}$ and for the signal cross section
without cuts, denoted $\sigma_{\rm signal}^{\rm no\;cuts}$, are shown
in the last two columns of Table~\ref{tab:angen}. Their
comparison shows  the pure off mass shell effects resulting from the
fact that the top quarks and Higgs boson are produced and they decay while
being off mass shell. 
%The pure off mass shell effects are also
%illustrated in Fig.~\ref{fig:offshell}.

In the next step, we define the following 
basic cuts which should allow to detect events with separate 
jets and/or isolated charged leptons:
\bea
5^{\circ} < \theta (q,\mathrm{beam}),\;\theta (l,\mathrm{beam})
 < 175^{\circ}, &\;\;\;&
\theta (q,q'), \;\theta (l,q) > 10^{\circ},\nonumber\\
E_{q},\;E_l,\;\; /\!\!\!\!E^T &>& 15\;{\rm GeV}
\label{cuts:semil}
\eea
and calculate the cross sections of (\ref{semileptonic}): 
$\sigma_{\rm all}$, with the complete set of Feynman diagrams, 
$\sigma_{\rm no\;QCD}$, without 
the gluon exchange diagrams and $\sigma_{\rm signal}$, with the 20 signal
diagrams of Fig.~\ref{fig:ee8f}.
To which extent cuts (\ref{cuts:semil}) reduce the signal can be
seen by comparing $\sigma_{\rm signal}$ with 
$\sigma_{\rm signal}^{\rm no\;cuts}$. A comparison of
$\sigma_{\rm all}$ and $\sigma_{\rm no\;QCD}$ with $\sigma_{\rm signal}$ shows
the full lowest order SM and pure EW off resonance background.
On the other hand, a comparison of $\sigma_{\rm all}$ with $\sigma_{\rm no\;QCD}$ 
illustrates the size of the pure QCD background which, in spite of what
could have been expected if taking into account law virtuality of
the exchanged gluons, turns out to be surprisingly large.
The cross sections $\sigma_{\rm all}$ calculated with {\tt carlomat} have been 
recalculated with {\tt O'Mega/Whizard}. A satisfactory agreement within 
$3\sigma$ for all reactions of (\ref{ee8f}) has been found \cite{KS1}.

At $\sqrt{s}=500$~GeV, where there is only very limited phase space volume 
for reaction (\ref{eetth}), the cross section depend substantially on
the actual values of $m_H$ and $m_t$.
This dependence for reaction 
(\ref{semileptonic}) is illustrated in Table~\ref{tab:massdep}.
\begin{table}
\begin{center}
\begin{tabular}{c|c|c|c|c|c}
\hline 
\hline 
\rule{0mm}{7mm} $\sqrt{s}$ & $\sigma_{\rm all}$ 
& $\sigma_{\rm no\;QCD}$  & $\sigma_{\rm signal}$ 
& $\sigma_{\rm signal}^{\rm no\;cuts}$
             & $\sigma_{\rm signal}^{\rm NWA}$\\
{\small [GeV]}& {\small [ab]} & {\small [ab]} & {\small [ab]} & {\small [ab]}
 & {\small [ab]} \\[1.5mm]
\hline 
\rule{0mm}{7mm}  
  500 &  26.8(4) & 7.80(3) & 3.095(3) & 3.796(3) & 3.920(1)  \\ [1.5mm]
  800 & 100.2(8) & 66.8(1) & 46.27(2) & 58.36(2) & 60.03(2)  \\ [1.5mm]
 1000 &  93.1(3) & 61.4(1) & 40.18(2) & 51.74(2) & 52.42(3) \\ [1.5mm]
 2000 &  47.4(2) & 28.5(1) & 15.14(3) & 22.14(4) & 20.68(3)
\end{tabular} 
\end{center}
\caption{Cross sections of reaction (\ref{semileptonic}) at different c.m.s.
energies with cuts (\ref{cuts:semil}) calculated: with the complete set of 
Feynman diagrams, $\sigma_{\rm all}$, without gluon exchange diagrams, 
$\sigma_{\rm no\;QCD}$, and with only the signal diagrams of 
Fig.~\ref{fig:ee8f}, $\sigma_{\rm signal}$.
The last two columns show the total signal cross section 
$\sigma_{\rm signal}^{\rm no\;cuts}$ and the total cross section in NWA
$\sigma_{\rm signal}^{\rm NWA}$ without cuts.}
\label{tab:angen}
\end{table}

\begin{table}
\begin{center}
\begin{tabular}{c|ccc|ccc}
\hline 
\hline 
\rule{0mm}{7mm}
&\multicolumn{3}{c|}{$m_H=115$~GeV}
&\multicolumn{3}{c}{$m_H=130$~GeV}\\[1.5mm]
\hline
$\;m_t$ &$\sigma_{\mathrm{all}}$ & $\sigma_{\mathrm{no\;QCD}}$
& $\sigma_{\mathrm{signal}}$ & $\sigma_{\mathrm{all}}$ 
& $\sigma_{\mathrm{no\;QCD}}$ & $\sigma_{\mathrm{signal}}$\\[0.1mm]
{\small [GeV]}& {\small [ab]} & {\small [ab]} & {\small [ab]} & {\small [ab]}
 & {\small [ab]}  & {\small [ab]} \\[1.5mm]
\hline
\rule{0mm}{5mm}
 171  & 39.8(1) & 20.21(5) & 14.61(1) & 29.9(1) & 10.29(3) 
& 4.670(4) \\
 174.3& 35.3(1) & 16.41(4) & 11.69(1) & 26.9(1) & 7.91(2) 
& 3.095(3) \\
 176  & 33.3(1) & 14.52(4) & 10.19(1) & 25.4(1) & 6.75(2) 
& 2.365(2)
\end{tabular} 
\end{center}
\caption{Lowest order cross sections of (\ref{semileptonic}) at
$\sqrt{s}=500$~GeV for different values of $m_H$ and $m_t$.}
\label{tab:massdep}
\end{table}

The off resonance background can be best reduced by imposing cuts on 
the invariant masses of two non $b$ jets
\bea
\label{nonb}
60\;{\rm GeV} < m\left(\sim b,\sim b'\right) < 90\;{\rm GeV},
\eea
and of a $b$ jet $(b_1)$ and the two non $b$ jets that have already 
passed (\ref{nonb})
\bea
\label{bnonb}
\left|m\left(b_1,\sim b,\sim b'\right)-m_t\right| 
            < 30\;{\rm GeV}.
\eea
In the semileptonic channels, we impose a cut on the transverse mass of the
lepton and missing energy system
\bea
\label{lmt}
m_T\left(l,\;/\!\!\!\!E^T\right) < 90\;{\rm GeV},
\eea
and then combine it with another $b$ jet $(b_2)$, and impose the cut
\bea
\label{blmt}
m_t-30\;{\rm GeV} < m_T\left(b_2,l,\;/\!\!\!\!E^T\right) 
< m_t+10\;{\rm GeV}.
\eea
See Eqs. (18) and (21) of \cite{KS} for the definitions of transverse masses in
(\ref{lmt}) and (\ref{blmt}).
Finally, we impose a cut on the other two $b$ jets, $(b_3\;{\rm and}\;b_4)$,
\bea
\label{bb}
\left|m\left(b_3,b_4\right)-m_H\right| < m_{bb}^{\rm cut}.
\eea
Cuts (\ref{nonb}) and (\ref{lmt}) should allow for identification of the 
$W$ bosons, cuts (\ref{bnonb}) and (\ref{blmt}) for identification 
of the top quarks 
and (\ref{bb}) for identification of the Higgs boson.
The cross sections of 3 different reactions representing
the hadronic, semileptonic and leptonic channels of (\ref{ee8f}) are shown 
in Table~\ref{tab:invmass}. We see that the background is 
substantially reduced.
\begin{table}
\begin{center}
\begin{tabular}{c|cc|cc|cc}
%\cline{2-7}
\hline
\hline
\rule{0mm}{6mm} 
$\epm$
 &\multicolumn{2}{c|}{$\ra b \bar b b \bar b u\bar d s \bar c$}
 & \multicolumn{2}{c|}{$\ra b \bar b b \bar b u\bar d \mu^- \bar \nu_{\mu}$}
 & \multicolumn{2}{c}{$\ra b \bar b b \bar b \tau^+ \nu_{\tau}
   \mu^- \bar \nu_{\mu}$}\\[0.2mm]
Cuts:
 &\multicolumn{2}{c|}{(\ref{nonb}), (\ref{bnonb}), (\ref{bb})}
 & \multicolumn{2}{c|}{(\ref{nonb})--(\ref{bb})}
 & \multicolumn{2}{c}{(\ref{bb})}\\[1.5mm]
\hline
\rule{0mm}{6mm} 
$m_{bb}^{\rm cut}$ & $\sigma_{\rm all}$  
& $\sigma_{\rm sig.}$  & $\sigma_{\rm all}$ 
& $\sigma_{\rm sig.}$  & $\sigma_{\rm all}$ 
& $\sigma_{\rm sig.}$  \\[0.2mm]
{\small [GeV]}& {\small [ab]} & {\small [ab]} & {\small [ab]} & {\small [ab]}
 & {\small [ab]}  & {\small [ab]} \\[1.5mm]
\hline 
\rule{0mm}{7mm}  
20 & 167.0(4)  & 128.4(1) & 43.6(1) & 33.93(2) & 23.28(5) & 13.48(1) \\ 
5  & 139.1(3)  & 128.0(1) & 35.8(1) & 33.10(2) & 16.95(4) & 13.47(1) \\ 
1  & 130.5(2)  & 127.7(1) & 33.4(1) & 32.82(2) & 14.44(4) & 13.46(1) 
\end{tabular} 
\end{center}
\caption{Lowest order cross sections of different
channels of (\ref{ee8f}) at $\sqrt{s}=800$~GeV with angular and
energy cuts (\ref{cuts:semil}) and the invariant mass cuts indicated above.}
\label{tab:invmass}
\end{table}
Let us note that a cut on the energy of a $b$ quark: $E_b > 40$~GeV 
or $E_b > 45$ GeV together with $m_{bb}^{\rm cut}=20$~GeV or $m_{bb}^{\rm cut}=5$~GeV
would also reduce the background, but it reduces the signal much more 
than cuts (\ref{nonb})--(\ref{blmt}) \cite{KS}.

\section{Summary and Outlook}

We have addressed the signal and background issue in the reaction 
of associated production of the top quark pair
and Higgs boson at the future $\epm$ linear collider that can be
used to determine the top--Higgs Yukawa coupling. 
The background contributions are large for typical
particle identification cuts. In particular, the QCD background is 
much bigger than it could have been expected taking into account 
a possibly low virtuality of exchanged gluons.
The background can be efficiently reduced by imposing invariant mass 
cuts allowing for the top and antitop quark, and Higgs boson identification. 
We have looked at the cross section dependence on the Higgs boson and top 
quark masses.
We have also illustrated pure off mass shell effects
by comparing the signal cross section calculated with the full 8 particle 
kinematics and in the NWA.
Taking into account the size of cross sections, the best place 
      to measure the top--Higgs Yukawa coupling seems to be a linear collider 
      operating at the centre of mass energy of about 800~GeV.

Acknowledgements: Work supported in part by the Polish Ministry 
of Science under Grant No. N N519 404034 and by European Community's 
Marie-Curie Research Training Network under contracts MRTN-CT-2006-035482 
(FLAVIAnet) and MRTN-CT-2006-035505 (HEPTOOLS).

\end{document}